\documentstyle[floats,prb,aps]{revtex}

\parskip=\medskipamount
\begin{document}
\input{epsf} 
\draft
\renewcommand{\textfraction}{0.0}
\renewcommand{\dblfloatpagefraction}{0.8}
\renewcommand{\topfraction}{1.0}   
\renewcommand{\bottomfraction}{1.0}   
\renewcommand{\floatpagefraction}{0.8}

\twocolumn[\hsize\textwidth\columnwidth\hsize\csname @twocolumnfalse\endcsname
								   
\title{Persistent currents in continuous one-dimensional
disordered rings within the Hartree--Fock approximation}
\author{A. Cohen$^{1,2}$, R. Berkovits$^{2}$ and A. Heinrich$^{1}$}
\address{$^1$Institut f\"ur Festk\"orper-und Werkstofforschung,
Dresden, 01171-Dresden, Germany \\
$^2$The Minerva center for the Physics of Mesoscopics, 
Fractals and Neural Networks,
Department of Physics, \\
Bar-Illan University, 52900 Ramat-Gan, Israel}
\date{\today}
\maketitle
\begin{abstract} 
{ 
We present numerical results for the zero temperature
persistent currents carried by interacting spinless electrons in
disordered one dimensional continuous rings.
The disorder potential is described
by a collection of $\delta$-functions at random locations and strengths.
The calculations are performed by a self-consistent Hartree-Fock (H-F)
approximation. 
Because the H-F approximation retains  the concept of 
single-electron levels, we compare the statistics of energy levels 
of noninteracting electrons with those of interacting electrons
as well as of the level persistent currents.
We find that the e-e interactions alters the levels and samples 
persistent currents and introduces a preffered diamagnetic current direction.  
In contrast to the analogous calculations that recently appeared in the 
literature for interacting spinless electrons in the presence of moderate 
disorder in tight-binding models we find no suppression of the 
persistent currents due to the e-e interactions.
}   
\end{abstract}


] 

\narrowtext 

\section{ INTRODUCTION }

The prediction
of B\"uttiker, Imry and Landauer \cite{bil} that a persistent
current (PC)\cite{by,blatt} can be observed in a disordered
normal metal mesoscopic
ring threaded by a magnetic flux, although this ring has a
finite resistance due to elastic scattering,
has attracted much recent attention
\cite{imry86,hofai88,helene89,gilles90,ashfy,schmeltz93}.
The elastic scattering, due to disorder, 
reduces the amplitude of this current.
The detection of the effect in three different experiments,   
by measuring the magnetic signal of the PC, has stimulated 
a great deal of theoretical interest in particular because 
of the large magnetic response that was measured.
The measured amplitude of the PC, although some of the
samples are clearly in the disordered diffusive regime, 
is of the same magnitude as that calculated  for
electrons in a clean ring.

The simultaneous measurement \cite{levy90} of the magnetic 
response of $10^7$ metallic Cu rings yields PC amplitude 
which is one order of magnitude larger than predicted by 
single-electron calculations which take into account 
the elastic scattering due to the disorder.
This rised the question of the source of such 
large magnetic response of an ensemble of rings 
which are in the diffusive regime of disorder. 
The experimental signals of three
single Au rings \cite{chand91} of different geometries 
were found to be up to two orders of magnitudes larger
than the prediction of such single electron theory.
On the other hand, the response \cite{mailly} of a 
2D semiconducting ring in the ballistic regime of disorder
and at a very low electronic density agrees well with the
theoretical predictions of the single electron theory.  
These experimental findings  
rise the question whether the discrepancies between
single electron theory and the observed behavior of the PC  
can be understood as a consequence of neglecting the e-e 
interactions in the presence of disorder.     
This question is part of a more general
problem of the influence of electronic interactions in the
presence of disorder on the properties of mesoscopic systems
which is in the center of many recent studies, such as polarizability
fluctuations \cite{ba}, quasiparticle life times in quantum dots
\cite{sia} and the metal-insulator transition \cite{belitz}.

There have been many attempts to investigate the influence
of e-e interactions on the PC.
Analytical calculations were mainly restricted to perturbation
theories\cite{eckschmiambe} and to renormalization 
treatments of Luttinger
liquid models \cite{models}, 
which have shown that some enhancement
of the PC due to the e-e interaction is possible.
Exact numerical diagonalization studies of spinless 
interacting electrons for
1D \cite{berkov,zerar} as well as for 2D \cite{ba1}
tight-binding (TB) models were performed.
In 1D systems for long range Coulomb interaction
the PC is weakly enhanced for strong disorder 
(i.e., for the localized regime)
and medium strength of interactions.
But for weak disorder the current is suppressed at any strength  
of the interaction and filling factor \cite{berkov}.
Also for any strength of short range e-e interactions, 
even far from half filling,  
the PC has shown a decrease\cite{zerar}.    
On the other hand, for 2D systems the PC is significantly enhanced
in the diffusive regime \cite{ba1}.
Although these studies are useful in pointing towards the 
general influence of the e-e interactions they are restricted, 
due to exponential expansion of the Hilbert space with sample size,  
to samples of small number of sites (about six to twenty).  
This complex behavior has its origin in the interplay 
between two different physical 
effects. 
While the disorder creates fluctuations in the electronic density 
leading eventually to localization, the e-e interaction is expected 
to reinforce the tendency for local charge neutrality which 
counteracts the influence of disorder. 
In the H-F approximation this is realized by the effective potential  
which is expected to lead to an increase in the PC amplitude.  
On the other hand the Mott-Hubbard transition\cite{mottliebwu} 
reduces the current amplitude because the e-e interaction 
opens a gap at the middle of the conduction band leading to an 
insulating --not metallic-- behavior of a half filled band.   

It is interesting to examine the PC behavior in the 
H-F approximation because exact analytical 
consideration of interaction in the presence of disorder 
is a very difficult task. Even perturbative calculations 
are not conclusive \cite{eckschmiambe}. 
On the other hand exact numerical diagonalization 
is limited to small number of sites and electrons. 
But using the H-F approximation has the advantage of  
the possibility to handle much larger systems. 
It has also  the possibility to differentiate between 
different influences of the e-e interactions which
may lead to an enhancement of the PC. 
Further it retains the concept of single 
electron levels in a meaningful way. 
This enables one to employ the considerable existing
knowledge on statistical properties of single electron energy 
levels in disordered systems \cite{mehta} for studying the influence
of e-e interactions. 
One is also able to study the behavior of the 
current for different regimes of disorder 
since as the single electron energy becomes higher 
the effective disorder for these energies is weaker.

Kato and Yoshioka \cite{yoshi} were the first to 
deal with large 1D TB 
system of 100 sites and about half filling in 
the frame work of the
self consistent H-F approximation for long range interaction.
They have found that for any value of disorder the 
electronic interactions further
suppress the PC and do not counteract the effect of disorder.  
This difference with the results of 1D exact 
diagonalization is due to the fact that the HF approximation 
for spinless electrons does not smear the density fluctuations  
which is essential to obtain an enhancement of the current in the
localized regime.
For spinless electrons in two\cite{zerar3d} and three\cite{yoshi1}
dimensional TB rings the HF calculations
show some enhancement of the average current for Coulomb interactions
in the diffusive regime.
But if the spin is taken into account the HF approximation
shows a large enhancement of  the PC\cite{zerar3da}.
The reason is that including the spin degrees of freedom 
adds a dominant contribution to the direct term.
In fact, this effect can be
seen even for first order perturbation theory\cite{bouchiat,asa}.
Thus, for the case where spin is taken into account, 
the HF approximation  does describe the 
appearance of local charge neutrality 
due to interactions.

The main goal of this paper is to compare the PC behavior 
of interacting spinless electrons 
in continuous disordered rings with that behavior in TB rings 
that recently appeared in the literature \cite{berkov,zerar,yoshi}. 
For these TB rings in the moderate regime of disorder it was found that 
e-e interaction further suppresses the PC because of the Mott-Hubbard 
transition which does not seem to be relevant for electrons in 
continuous rings. 
The difference in the behavior of the PC is a measure for the 
importance of the difference between continuous and discrete models. 
The difference between a model with no discrete symmetries and 
the TB model has been 
suggested in Ref.\onlinecite{weiden} and here we shall 
examine this difference for a well defined Hamiltonian.    

The rest of the paper is organized as follows:
in section II we present the model for independent electrons and 
interacting electrons within the self consistent HF approximation.
In section III we discuss the numerical results,
and in section IV we derive our conclusions.

\section{Model of electrons in disordered 1D continuous rings.}

\subsection{Non interacting electrons}

Let $R$ be the radius of the continuous 1D ring,
$\theta$ the coordinate along the ring,
$m_e$ is the electron mass, $e$ its charge,
$\phi$ the flux threading the ring and
$\phi_0 \equiv {hc \over e}$ the flux quantum.
For completeness and further reference let us 
restate some basic facts. 
Defining the disorder potential as
$V_{dis}(\theta)$ 
the Schr\"odinger equation 
(in energy units ${\hbar^2 \over m_e R^2}=1$ )   
for non-interacting
electrons is:
\begin{equation}
{1\over 2} \left(  { \partial \over i \partial \theta} +
{\phi \over \phi_0} \right)^2 \Psi(\theta) +
V_{dis}(\theta) \Psi(\theta) = E \Psi(\theta) ,
\label{sch0}
\end{equation}
where the boundary conditions (BC)  are periodic
\begin{equation}
\Psi(\theta+2\pi)=\Psi(\theta).
\label{bc0}
\end{equation}
The substitution\cite{by}
$\; \Psi(\theta) = \psi(\theta) e^{-i{\phi \over \phi_0}\theta} \;$
yields a wave equation of the Bloch type\cite{bil}
\begin{equation}
-{1 \over 2} { \partial^2 \over \partial \theta^2} \psi(\theta) +
V_{dis}(\theta) \psi(\theta) = E \psi(\theta)   ,
\label{sch1}
\end{equation}
which now has the modified BC
\begin{equation}
\psi(\theta+2\pi) = \exp{(i2\pi{\phi \over \phi_0})} \psi(\theta).
\label{bc1}
\end{equation}
An exact eigenstate of energy $E_l$
carries PC $i_l$   
\begin{equation}
\begin{array} {c}
i_l=-c{\partial \over \partial \phi}E_l
\equiv
-{\hbar \over m_e R} <\Psi_l|
\left( {\partial \over i\partial \theta} +
{\phi \over \phi_0} \right)|\Psi_l> \\ =
-{\hbar \over m_e R} <\psi_l|
{\partial \over i\partial \theta}
|\psi_l>      
\end{array}
\label{exacti}
\end{equation}
(where numerically we use dimensionless energy and dimensionless current
$ - {\partial E \over \partial {\phi \over  \phi_0 } } $). 
We stress that $\Psi_l(\theta)$ are the electronic wave functions
which are exactly periodic for any value of $\phi$. This fact will be
important when we consider interactions.
For $V_{dis}(\theta) \equiv 0$ the electronic functions are
$\Psi_l(\theta)={1\over \sqrt{2\pi}} \exp{(i l \theta)}$
with
$l=0,\pm 1, \pm 2 ,...$
and the spectrum is
$ E_l = {1\over 2} x_l^2$ with $x_l\equiv l+ {\phi \over \phi_0}.$
For our model we have chosen
\begin{equation}
V_{dis}(\theta)=\sum_{j=1}^{N_s} \lambda_j \delta(\theta-\theta_j)
\label{vdis}
\end{equation}
where the locations $0\le \theta_j \le 2\pi$ and strength
$-\Lambda \le \lambda_j \le \Lambda$ of the individual $j$-th
scatterer  are random with the appropriate
probabilities: $P_{_\theta}(\theta_j)={1\over 2\pi}$ and
$P_{_\lambda}(\lambda_j)={1\over 2\Lambda}$.
$N_s$ is the total number of scutterers in the ring.
$V_{dis}(\theta)$ produces characteristics of
disordered samples as was discussed by Imry and Shiren\cite{shiren}
while truncating the Hilbert space.
For non-interacting electrons one can {\em exactly}
find the spectrum and eigen functions by
means of a numerical transfer matrix technique.
A transfer matrix that ``propagates'' the solution of eq.(\ref{sch1})
from the left of the $j$-th scatterer
to the left of the next $j+1$-st scatterer is given by:
\begin{equation}
{\bf T}_j =  \left(
\begin{array}{cc}
(1+{\lambda_j \over i x}) e^{ix(\theta_{j+1} -\theta_j)} &
{\lambda_j \over i x}     e^{ix(\theta_{j+1} -\theta_j)}\\
-{\lambda_j \over i x}    e^{-ix(\theta_{j+1}-\theta_j)} &
(1-{\lambda_j \over i x}) e^{-ix(\theta_{j+1}-\theta_j)}
\end{array}
\right)
\label{tmatrix}
\end{equation}
where the exponents contain the effect of free propagation
of waves between the scutterers.
If we denote as
${\bf v}_{j}$ a two component vector in which
the upper component corresponds to the amplitude of the
forward propagating wave $e^{ix\theta}$ and the lower
component to $e^{-ix\theta}$ then ${\bf v}_{j+1}={\bf T}_j {\bf v}_j$.

The concepts of localization length $\xi$ and 
the Landauer conductance, which
are very useful in the study of open systems,
can be defined by comparing eq.(\ref{tmatrix}) with the general form
of a transfer matrix:
\begin{equation}
{\bf T} = \left(
\begin{array}{cc}
1/t^{*} & -(r/t)^{*} \\
-(r/t)  & 1/t
\end{array}
\right)  .
\label{tmatform}
\end{equation}
($t$ and $r$ the complex transmission
and reflection coefficients obeying $|r|^2+|t|^2=1$).
One can now analytically find $\xi$ as a
function of energy and also find the appropriate
scaling\cite{scaling} behavior of the
fluctuations of the Landauer resistance
$\rho \equiv {|r|^2 \over |t|^2}$
(or conductance $g\equiv 1/\rho$), where $N_s$
plays the role of the length scale of the conductor\cite{lscaling}.
\begin{figure}[b]
\centerline{\epsfxsize = 9cm \epsffile{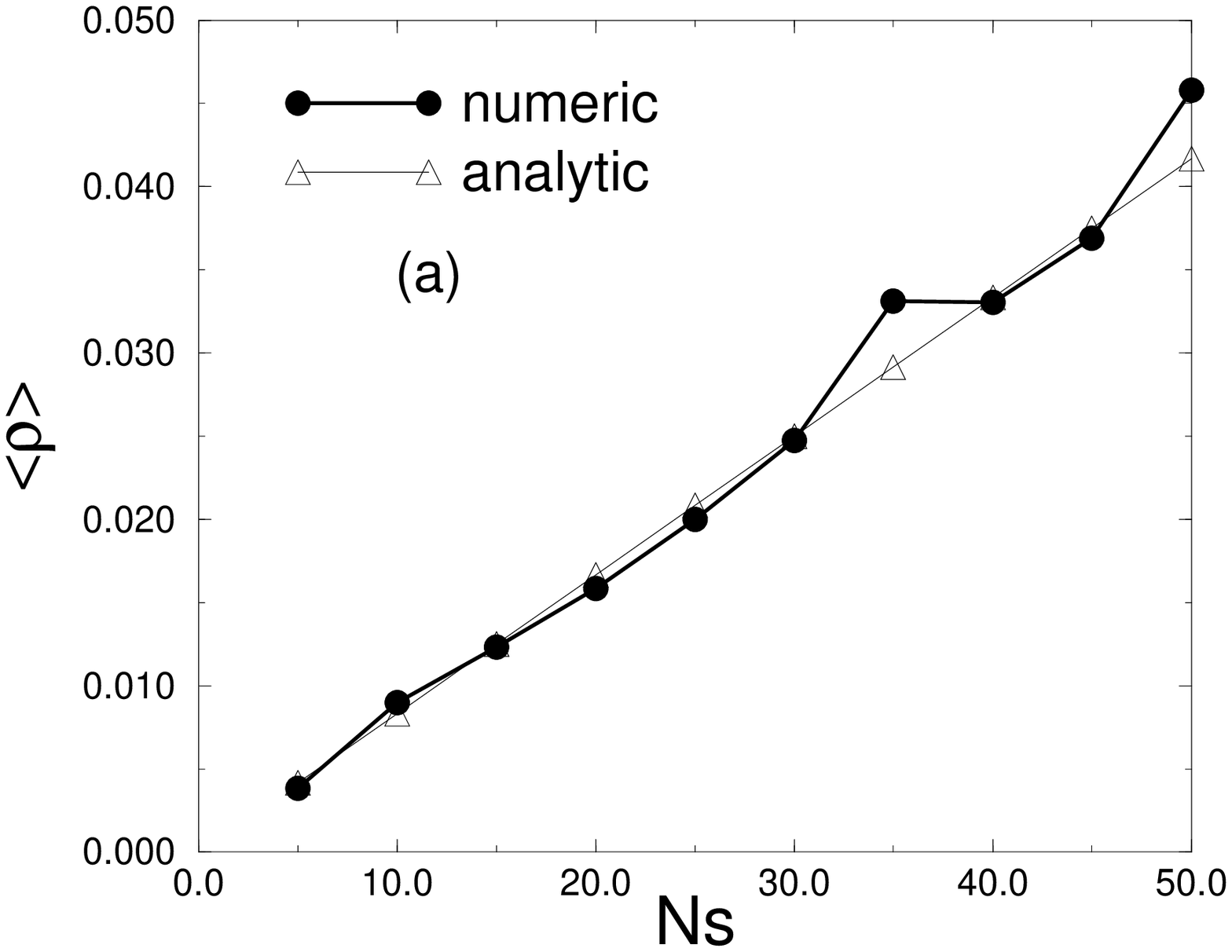}}
\centerline{\epsfxsize = 9cm \epsffile{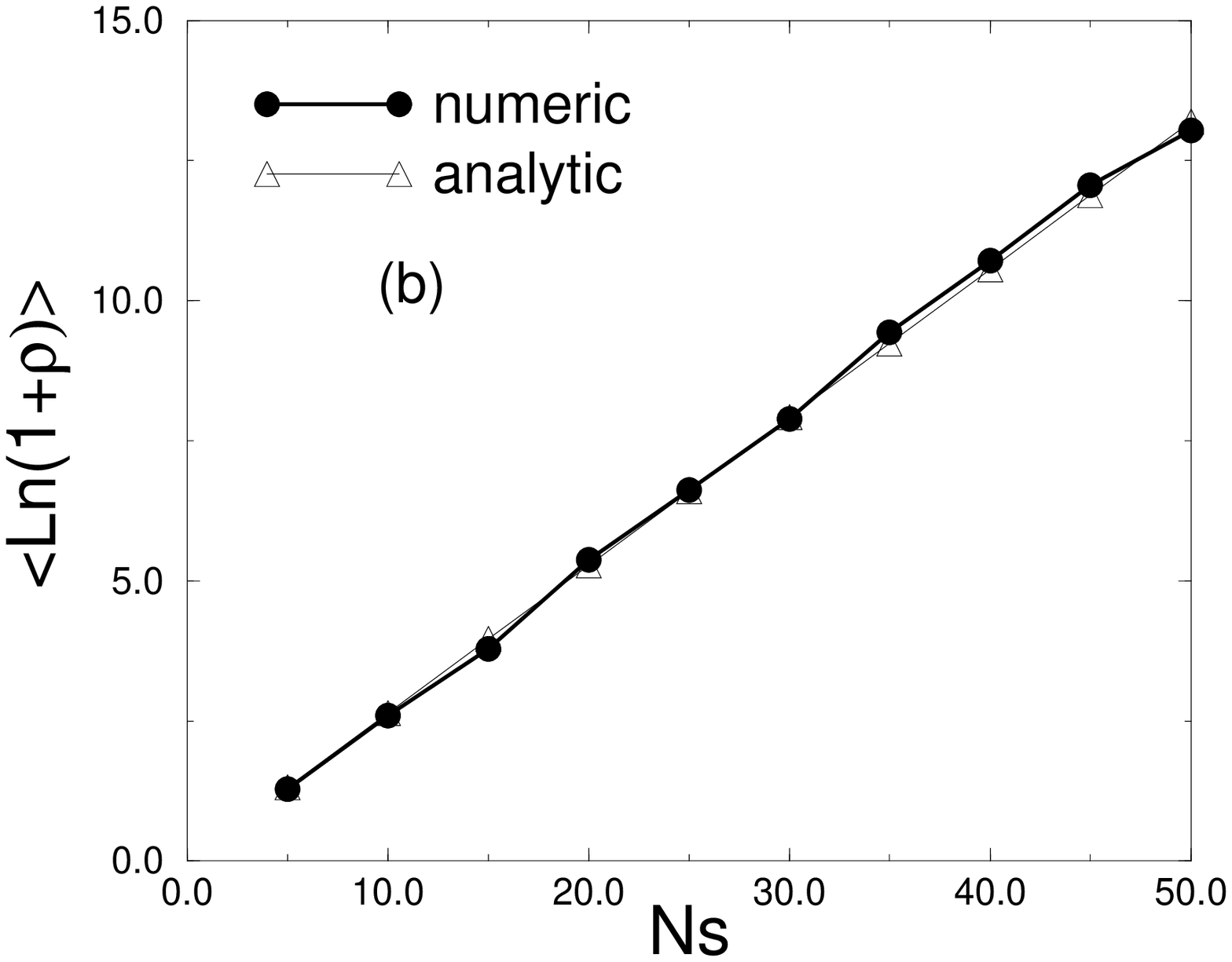}}
\caption{
(a) Scaling behavior of the disorder average Landauer 
resistance $<\rho>$
as a function of the number of scatterers $N_s$ for
an open chain of scatterers at weak disorder ($\Lambda=5$, $E=5000$,
see text).
(b)
Scaling behavior of $<\ln(1+\rho)>$ of an open chain at strong disorder
($\Lambda=20$, $E=200$).
The slope defines the inverse localization length ${1\over \xi}$.
The empty symbols in (a) and (b) are our analytical estimates by
eqs.(\protect{\ref{avrho}}) and (\protect{\ref{lnrho}}) respectively,
the full symbols are results of numerical 
simulations of 200 realizations.
}
\label{fig1}
\end{figure}
For weak disorder, where  
$\rho_j \equiv {x^2 \over \lambda_j^2}<< 1,$
and small $N_s$ one expects an Ohm law behavior,
i.e., disorder averaging should
give $<\rho > \sim <\rho_1> N_s$
($<\rho_1>$ is the average Landauer
resistance of an individual scatterer).    
One finds 
\begin{equation}
<\rho > \sim {\Lambda^2 \over 3x^2}N_s ,
\label{avrho}
\end{equation}
as long as $<\rho >$ is of order unity or smaller.
A comparison of numerical results
with the above equation is presented in Fig.\ref{fig1}a.

Because of localization,
for strong disorder or at any large enough $N_s$ (where $1<< \;<\rho>$)
we expect
$<\ln(1+\rho)> \sim {1 \over \xi}N_s$ .
In our case we analytically calculate
$\xi$ to be
\begin{equation}
\xi^{-1}\equiv <\ln (1+ \rho_1)>=
\ln (1+{\tilde \Lambda}^2) - 2 +
{2\over {\tilde \Lambda}} \tan ^{-1}{\tilde \Lambda}
\label{lnrho}
\end{equation}
where ${\tilde \Lambda} \equiv {\Lambda\over x}$ and $\rho_1$ is
the resistance of an individual scatterer.
This result is confirmed by numerical calculations presented in
Fig.\ref{fig1}b where the slope is the inverse localization length.

Let us now return to the discussion of PC in the closed ring.
The discrete spectrum and eigen functions of the ring can be found
using the definition of the transfer matrix (eq. (\ref{tmatrix}))
and the BC (eq. (\ref{bc1})) as an eigenvalue equation:
\begin{equation}
\prod_{j=1}^{N_s}  {\bf T}_j {\bf v}_1 =
e^{2\pi i{\phi\over\phi_0}}{\bf v}_1  .
\label{spectrum}
\end{equation}
It is enough to deal with the real (or imaginary) part of the
last eigenvalue condition\cite{lee80cohen95} in order to obtain
the eigen values because the other part is automatically fulfilled.
The case of $N_s=1$ at any strength of $\lambda$ can easily be
solved.
For $1<<N_s$ and appreciable strength of $\lambda_j$ the current
can be calculated numerically and is found to be  
suppressed by many orders of magnitudes due
to localization.

For a given disorder the total current\cite{altshuler91} 
in an isolated (canonical) sample
is the sum of the currents of all the occupied levels in the ring, 
which for $N_e$ electrons at zero temperature corresponds to:
\begin{equation}
I_{c}=\sum_{l=1}^{N_e} i_l .
\title{ic}
\end{equation}
For a sample connected to a reservoir (grand canonical)
the sum is over all levels having
energies lower than a given chemical potential $\mu$:
\begin{equation}
I_{g}=\sum_{E_l \le \mu} i_l .
\label{ig}
\end{equation}

\subsection{Interacting electrons}

We include $e-e$ interaction within the self 
consistent HF approximation.
In analogy to the notations in eqs.(\ref{sch0}) -- (\ref{bc1}) our
equation for the {\em new} HF single particle wave function
$\psi(\theta)$ reads:
\begin{equation}
\begin{array}{l}
-\; {1 \over 2} { \partial^2 \over \partial \theta^2} \psi(\theta) +
V_{dis}(\theta) \psi(\theta) \\
\; \\ 
+\;
{R\over r_0}
[
{-N_e  \over 2 \pi} \int_0^{2\pi}
{ d\theta' \over \sqrt{(\theta-\theta')^2+\epsilon^2}  } 
\psi(\theta)  \\
\\ 
+\;
\int_0^{2\pi}  { \sum_{l=1}^{Ne} | \Psi_l(\theta') |^2
\over \sqrt{(\theta-\theta')^2+\epsilon^2}   } 
d{\theta'} \psi(\theta) \\
\; \\ 
-\;
\int_0^{2\pi}
{ \sum_{l=1}^{Ne}\Psi_l^{*}(\theta')\Psi_l(\theta)
\over \sqrt{(\theta-\theta')^2+\epsilon^2}   }
\psi(\theta')d\theta'
]
= E \psi(\theta).
\end{array}
\label{hfintegro}
\end{equation}
with the BC of eq.(\ref{bc1}).
$r_0 \equiv {\bf \varepsilon} \hbar^2/m_e e^2$,
and ${\bf \varepsilon}$ is the dielectric constant,
to be distinguished from the cutoff $\epsilon$.
The first term in the above square brackets takes into account
the effect of the neutralizing background
charge. 
This results in a constant contribution to the potential which
will not affect the PC. The second term corresponds to
the Hartree (direct) term and the third contribution
represents the Fock (exchange) term.
The square distance between the particles is approximated by
$(\theta-\theta')^2 \equiv
\min [|\theta-\theta'|^2, (2\pi-|\theta-\theta'|)^2]$.
A cutoff $\epsilon^2$ was introduced in order
to make the contribution of each term finite.
If one ignores the background term
there is no need to introduce a cutoff since
the divergence of the direct and exchange terms will 
cancel each other.
Nevertheless, for numerical convenience, 
we treat each term separately
and check that this cutoff has no influence on the results.

In order to approximate the integro-differential equation 
(Eq.(\ref{hfintegro}))
by an ordinary differential equation
we would like to approximate the exchange term by a term which
can be represented as an effective potential.
This may be achieved by using the almost closure relation,
$\sum_{l=1}^{Ne}
\Psi_l^{*}(\theta')\Psi_l(\theta)
\sim \delta (\theta'-\theta)$,
in the Fock term, which is reasonable
when $N_e\gg1$. The last sum has a finite width which is of 
order ${1 \over N_e}$ 
This means that the main contribution to the Fock term integral
comes from $\theta' \sim \theta$. Therefore
one can exclude the unknown function
$\psi(\theta')$ from the integrand and replace it by
$\psi(\theta)$ as a multiplicative factor\cite{uslandau}.
Thus in this approximation the Fock term is replaced by
$
\sim {R\over r_0} \int_0^{2\pi}
{ \sum_{l=1}^{Ne}\Psi_l^{*}(\theta')\Psi_l(\theta)
\over \sqrt{(\theta-\theta')^2+\epsilon^2}   }
d\theta'\psi(\theta)
$
and eq.(\ref{hfintegro}) becomes an ordinary Schr\"odinger equation:

\begin{equation}
-{1 \over 2} { \partial^2 \over \partial \theta^2}
\psi(\theta) +
V_{dis}(\theta) \psi(\theta) +
{R\over r_0} V_{eff}(\theta) \psi(\theta)
= E \psi(\theta).
\label{hfeq}
\end{equation}
\begin{figure}[t]
\centerline{\epsfxsize = 9cm \epsffile{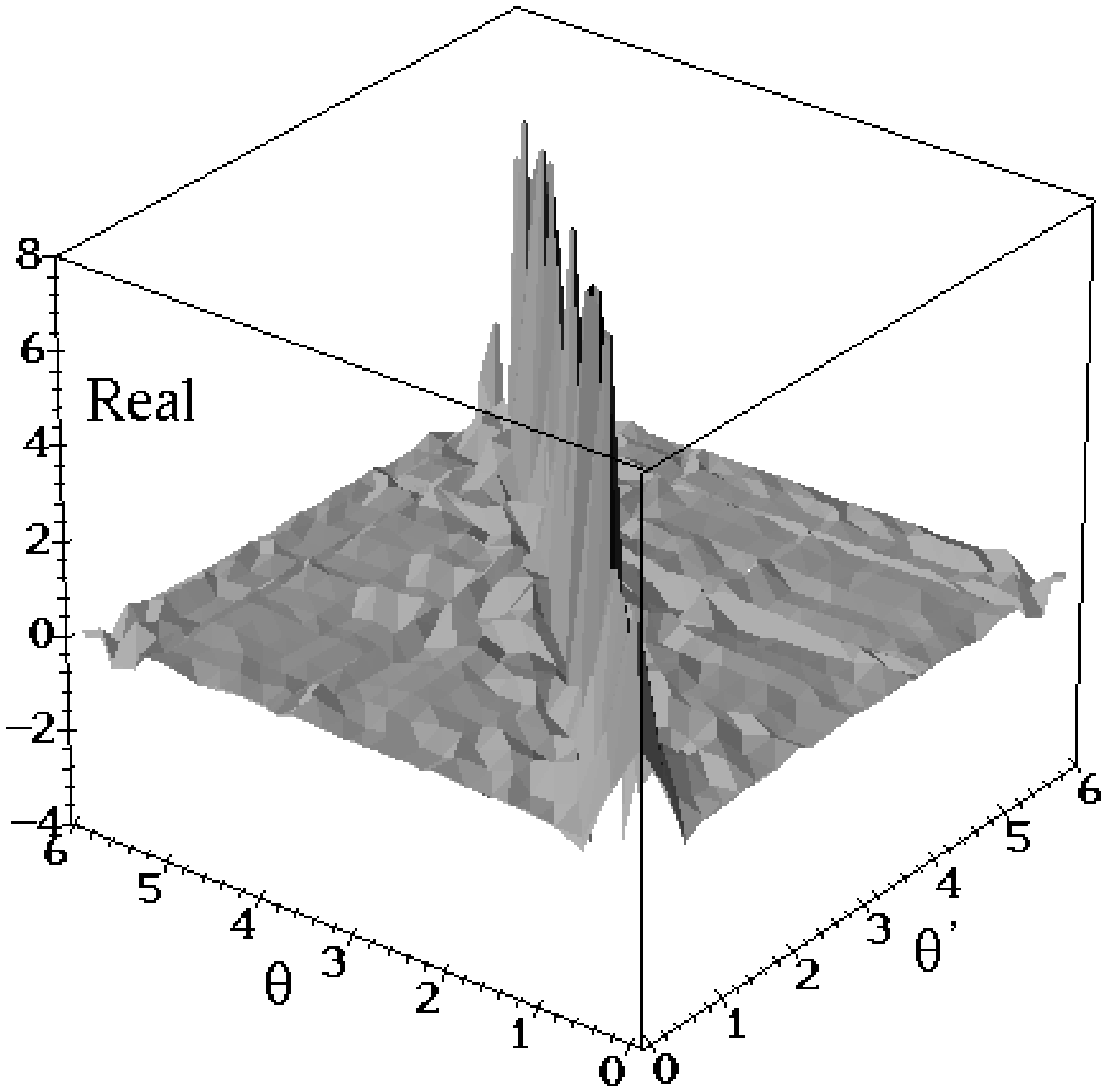}}
\centerline{\epsfxsize = 9cm \epsffile{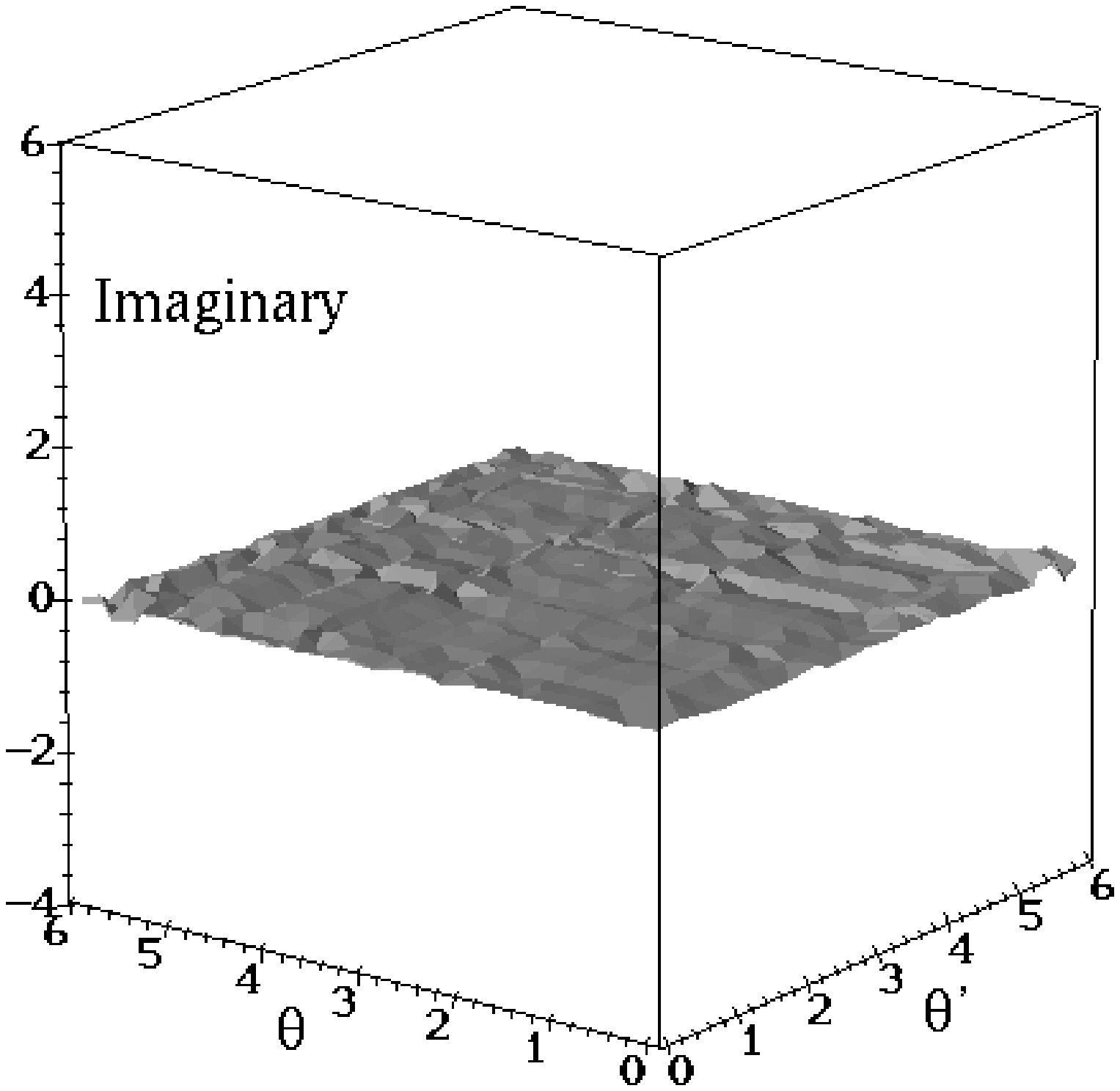}}
\caption{
Presentation of the almost closure relation:
$\sum_{l=1}^{Ne}  \Psi_l^{*}(\theta')\Psi_l(\theta)
\sim \delta (\theta'-\theta)$,
see eqs.(\protect{\ref{hfintegro}}) and (\protect{\ref{hfeq}}),
for one ring of the disordered ensemble
at convergence of the self consistency
($N_s=30$ random scatterers, $\Lambda=14$, $Ne=42$ electrons).
The real part of the sum is concentrated at
$\theta \sim \theta'$.
The imaginary part of the sum is vanishingly small.
This figure justifies the approximation of the HF 
integro differential
eq.(\protect{\ref{hfintegro}})  by  eq.(\protect{\ref{hfeq}}) that
we self consistently solved.
Note that the electronic wave functions $\Psi_l(\theta)$ are $2\pi$
periodic for any flux and so is the effective potential.
}
\label{fig2}
\end{figure}
Here $V_{eff}(\theta)\psi(\theta)$ contains the three terms in the
square brackets of eq.(\ref{hfintegro}) with the approximated Fock
term. The result of  $\sum_{l=1}^{Ne}
\Psi_l^{*}(\theta')\Psi_l(\theta)$ for a typical realization
is presented in Fig.\ref{fig2}. It is clearly seen that,
due to the sharpness of the almost closure relation,
approximating eq.(\ref{hfintegro}) by eq.(\ref{hfeq}) is 
reasonable.  
The approximation is reasonable for all electrons. 
At low energies the wave length of the considered electron 
is much larger than the width ${1 \over N_e} \sim \lambda_{F}$ 
(where $\lambda_{F}$ is the Fermi wave length) and therefore 
the approximation is reasonable at low energies.  
For an electron near the Fermi energy $E >> V_{eff}(\theta)$ 
(where $V_{eff}(\theta)$ is defined to vanish 
at its minimum and is of the order of $\Lambda$)   
the error as a consequence of the approximation can not create 
any significant change in the current of such electron.  

We solve eq.(\ref{hfeq}) self-consistently. For the n-th 
iteration the single electron wave functions from the previous 
iteration are taken and
used to construct the effective potential
$V_{eff}(\theta)$ for the current iteration.
For the first iteration ($n=1$) we define $V_{eff}(\theta)$ by the
non-interacting solutions,
i.e., equivalent to setting 
$V_{eff}(\theta) \equiv 0$ at iteration $n=0$.
The smooth $V_{eff}(\theta)$ is approximated by a constant potential
in the intervals between the external impurities. If $N_s$ is small
a finer subdivision of the intervals may be necessary.
For the k-th interval $\theta_k \le
\theta < \theta_{k+1}$ the effective potential is defined as
$V_{eff}(\theta)=V_{eff}(\theta_k)$.
The orthogonal
solutions, which vary from one interval to the other,
are given by $e^{\pm z_k \theta}$
where $z_k\equiv \sqrt{2(V_{eff}(\theta) -E)} $ is either real or
pure imaginary.
As one is free to add a global constant to the energy we define
$V_{eff}(\theta)$ to be zero at its minimum.
For that interval the general  
solution $\psi(\theta)$ is a linear combination of
$\exp(\pm i x\theta)$ with some real positive $x$ that defines the
energy $E=x^2/2$.
Propagating $\psi(\theta)$ along the ring circumference\cite{atvmin},
with an appropriate
matching at the boundaries of all intervals, and applying
the BC of eq.(\ref{bc1})  yield the discrete spectrum
\cite{searchx} and eigen functions of eq.(\ref{hfeq}) which
can now be normalized.

We shall now discuss the choice of parameters for the numerical
solution of Eq.(\ref{hfeq}).
The number of electrons in a given realization
of disorder was set by requiring that the energy 
of the highest occupied state
$l$ will be smaller than $312.5$, 
i.e., $x_l<25$, for any value of flux.
This results in an average filling of $42$ electrons per realization.
This requirement was chosen in order to 
balance between computation time
limitations and maintaining enough electrons
for the validity of the almost closure relation. 
This choice allows
the study of different energy regimes such as the almost 
ballistic (higher energies),
moderate (intermediate energies) and the strong (lower energies)
disorder limit. 
The strength of the interaction term is determined by the
ratio ${R\over r_0}$. For metals this ratio for a reasonable
experimental setting (say $R=0.1 \mu m$) will be about $10^3$, while
for semiconductors (which have small effective
masses and large
dielectric constants) the ratio is of order
of one. Since we can treat only a limited number of electrons, we
are really considering the low density limit of the problem, which
corresponds better to the semiconductor case and therefore we chose
${R\over r_0}=1$. 
A comparison between the TB model parameters and our estimation
of the strength of interaction in semiconducting rings can be achieved
by recognizing that for the typical densities of semiconducting devices 
the ratio of the electrostatic interaction 
energy $U$ to the kineic energy $t$ (the hopping matrix element)  
is of order of unity\cite{semicond} for semiconductor. 
The cutoff was chosen as $\epsilon=10^{-5}$.
It was checked that enlarging the cutoff up to $10^{-3}$
does not change the results within the limits of accuracy.
As a convergence condition for the $n$-th iteration at a given
flux we required:
$ \sum_{l=1}^{Ne} \mid x_l^{(n)} - x_l^{(n-1)} \mid < 0.00025 $.
For a typical realization this condition was fulfilled
within ten iterations.

In order to calculate the PC one can use two different approaches.
The first is the derivative of the many particle ground state
energy as function of the
flux and the second is a direct application of the current operator
on the many particle wave function. As we have confirmed both 
methods should give the same current,
i.e.,
\begin{equation}
I_v=
-c{\partial \over \partial \phi}  <\Psi_v|{\bf H} |\Psi_v> =
<\Psi_v|{\bf \hat J} | \Psi_v>.
\label{iv}
\end{equation}
Here, $I_v$ and $\Psi_v$ are the H-F ground PC and H-F ground state,  
${\bf H}$ is the exact Hamiltonian. 
In the calculation of the current it is sometimes convenient to
use the following relation\cite{merzbacher}
\begin{equation}
\begin{array}{c}
-c{\partial \over \partial \phi}  <\Psi_v|{\bf H} |\Psi_v> \\ =
-c\sum_{l=1}^{N_e}
\left[
{\partial \over \partial\phi}E_l -
<\psi_l|\left({\partial\over \partial\phi}{\bf V}_{eff}\right)|\psi_l>
\right]
\end{array}
\label{devdphi1}
\end{equation}
where $E_l$ are the eigenvalues of eq.(\ref{hfeq}). It is worthwhile
to note that in the H-F approximation the PC is not simply the
derivative of the sum of the H-F single particle eigenvalues $E_l$.
The single level current for a particular level is defined
as $i_l = -c \left[
{\partial \over \partial\phi}E_l -
<\psi_l|\left({\partial\over \partial\phi}{\bf V}_{eff}\right)|\psi_l>
\right]$.
We shall note that the PC and the eigen states of a clean ring are completely 
unaffected by the e-e interaction.

\section{Numerical results}

In this section we shall describe the results of our self consistent
numerical solution of Eq. (\ref{hfeq}).
We have considered realizations with $N_s=30$ randomly placed
delta scatterers of strength $\Lambda = 14$ (see eq.(\ref{vdis}))
for all the numerical calculations.
The number of electrons occupying a specific realization was defined 
by the noninteracting problem as was explained in the previous sub-section.
For a typical realization $N_e$ was of order of $42 \pm 6 $.
$V_{eff}(\theta)$ was calculated
as a staircase potential
at $2N_s$ points (see the previous section for details)
and was found to be  a reasonable
approximation of the smooth effective potential.
Our ensemble of disordered rings contained 150 different realizations.
In the following we present the results of our study of the effect 
of the e-e interactions on (A) the level spacing, (B) the PC distributions,  
and (C) the canonical average and the single ring PC.   
					 
\subsection{Spectrum}
\begin{figure}[ht]   
\centerline{\epsfxsize = 9cm \epsffile{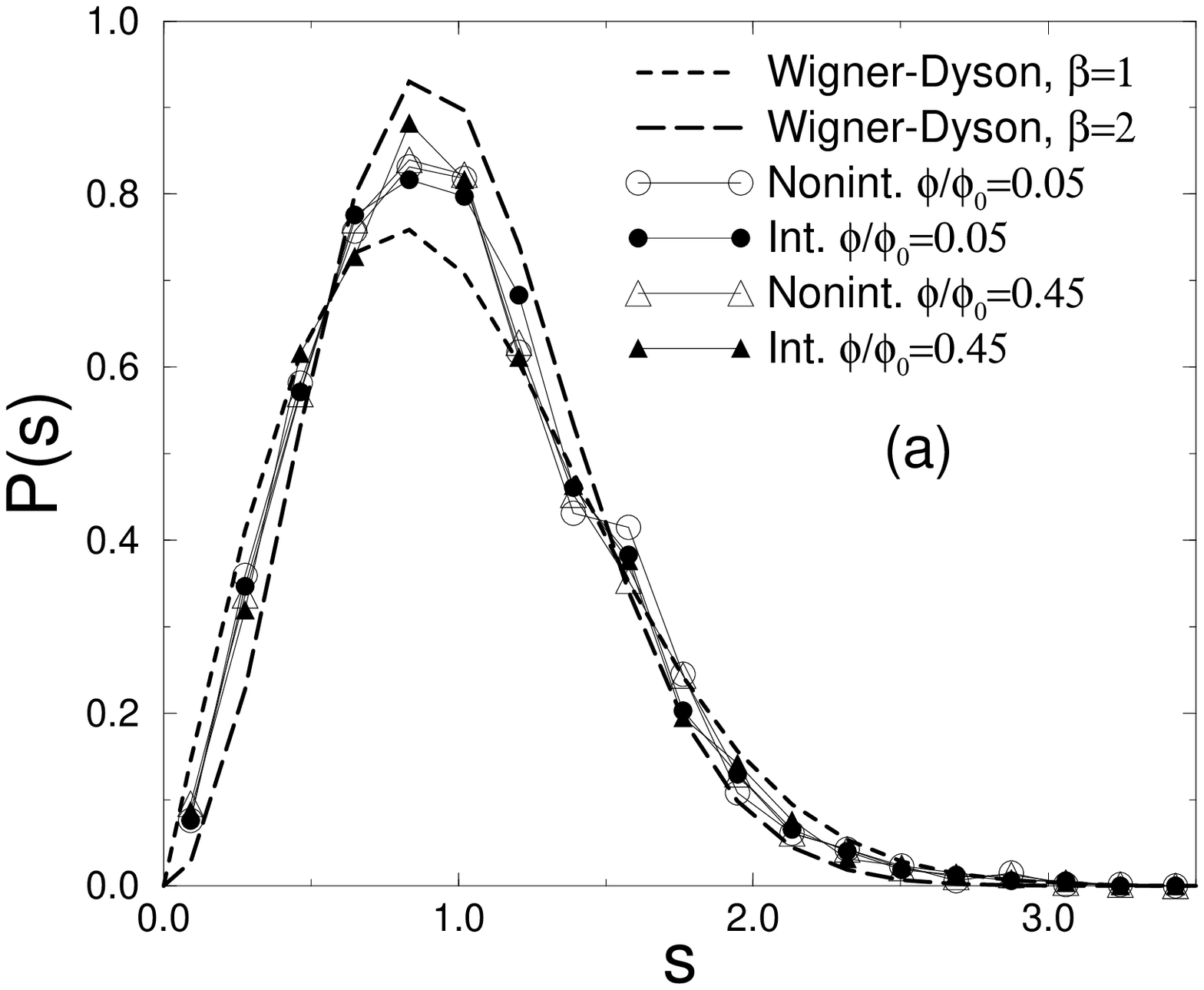}}
\centerline{\epsfxsize = 9cm \epsffile{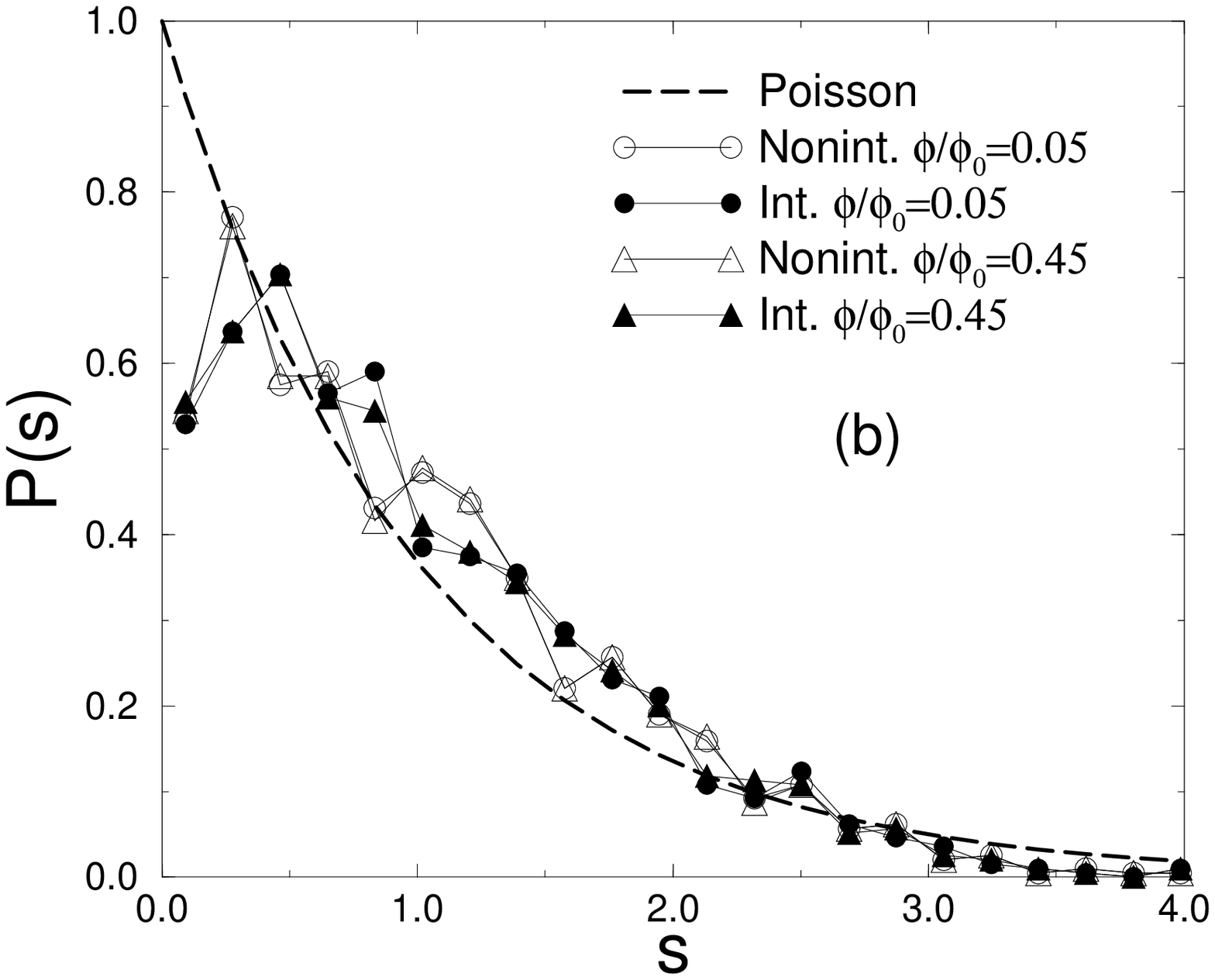}}
\caption{
Comparison of level spacing statistics $P(s)$
of noninteracting (empty symbols) and interacting
(bold symbols) electrons within the H-F approximation.
(a) At high energy strip $P(s)$ is Wigner-Dyson
$N(\beta)s^\beta\exp(a_\beta s^2)$.
$\beta=1, \; a_1=\pi/4, \; N(1)=\pi/2$ for zero flux (dashed),
$\beta=2, \; a_2=4/\pi \; N(2)=32/\pi)$ for half quantum of flux
(long-dashed).
(b) At low energy strip the states are localized also when including
interaction as can be seen from the statistics 
which remains Poissonian.
Because the change from (a) to (b) is 
continuous as a function of energy
and $P(s)$ is unchanged when including 
interaction there is no sign of
interaction induced insulator to metal transition.
}
\label{fig3}
\end{figure}

As mentioned in the introduction, one of the main advantages of the
H-F approximation is that although one is dealing with a many particle
problem the single electron levels still characterize the system.
Thus, the immediate question arises whether e-e interactions
imposes a transition on the level statistics.
Because in the H-F approximation 
the system is still characterized by a single electron
spectrum of an effective Hamiltonian (see eq.(\ref{hfeq})),
we were trying to answer 
this question by comparing the level spacing statistics of the 
eigenvalues of eq.(\ref{sch1}) with
that of eq.(\ref{hfeq}) when the condition for self consistency 
is satisfied. 
The level spacing statistics are considered
for different regimes of disorder:
(i) the extended regime (levels within a high energy strip) which  
corresponds to Wigner statistics for the noninteracting electrons; 
(ii) the localized regime (levels within a low energy strip)   
corresponding to the Poisson statistics for the
noninteracting electrons. 
For these low energies the levels are clearly localized 
because the calculated levels currents were found to be 
exponentially small and 
by drawing  
$| \psi_l (\theta) |^2$ we simply visualized the localization.  
Fig.\ref{fig3} shows the results for both regimes. 
There is no transition in the level spacing due 
to the e-e interactions. 
The same analysis was carried out for an intermediate regime 
(levels in a regime between (i) and (ii))  
also did not show any    
difference between the statistics of the interacting and 
noninteracting electrons, although in both of these cases  
the expected transition
from Poisson to Wigner statistics as a function of energy 
was found.  
Thus, the level statistics do not show any sign  
of an insulator to metal transition, which may be 
expected if the interactions
induce a straight forward change in the localization properties
of the H-F levels. 
Therefore, if there would be any enhancement of the PC 
(of this spinless model) it 
can not be explained  
by an interaction induced insulator to metal transition.

\subsection{Distributions of PC}

\begin{figure}[b]
\centerline{\epsfxsize = 9cm \epsffile{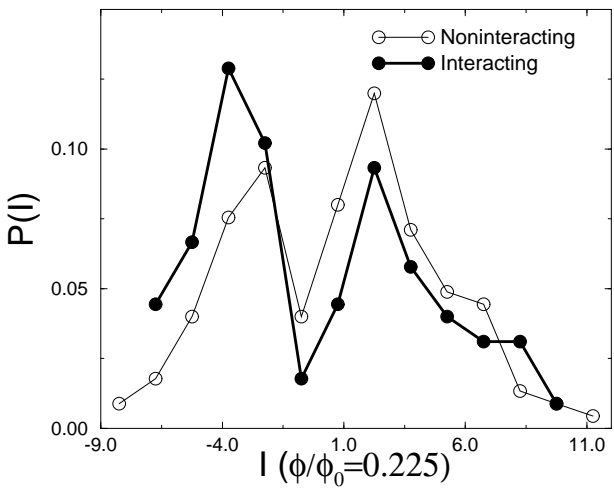}}
\caption{
Probabilities of sample PC for interacting (bold symbols) and 
noninteracting
(empty symbols) electrons in the {\em same} canonical ensemble
of 150 disordered rings
at ${\phi \over \phi_0} =0.225$.
}
\label{fig4}
\end{figure}
 
In order to understand the effect of e-e 
interactions on the canonical PC
we plot the distribution of the canonical persistent currents for
an ensemble of realizations with
non-interacting electron as well as for the same
realizations when interactions are taken into account.
As can be clearly seen in Fig.\ref{fig4} the
distribution of sample current is shifted towards more
negative values,
i.e. the PC becomes diamagnetic.
Nevertheless, as a result of the interactions,
individual realizations can change the current size 
and current direction in both the paramagnetic and
diamagnetic directions.
Unfortunately, because our ensemble is small it is 
meaningless to
define the enhancement factor of the typical 
$\sqrt{\langle I^2 \rangle}$ PC
since it is strongly controlled by rare events in the tails.
This shift of the distribution is similar to the situation seen in
an exact diagonalization study of small 2D systems in the diffusive 
regime \cite{ba1}, although the direction is opposite\cite{bouchiat}.

\begin{figure}[b]
\centerline{\epsfxsize = 9cm \epsffile{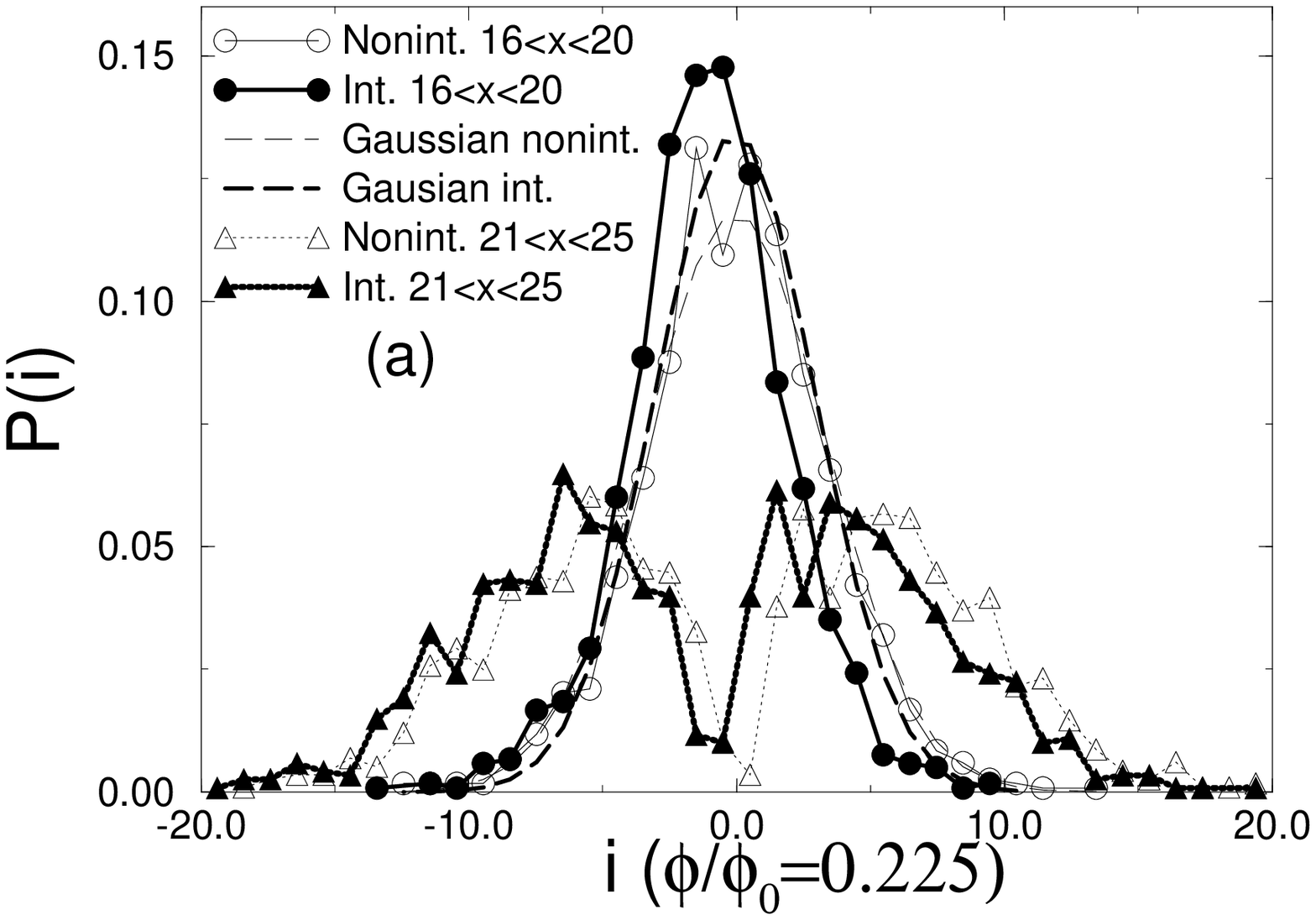}}
\centerline{\epsfxsize = 9cm \epsffile{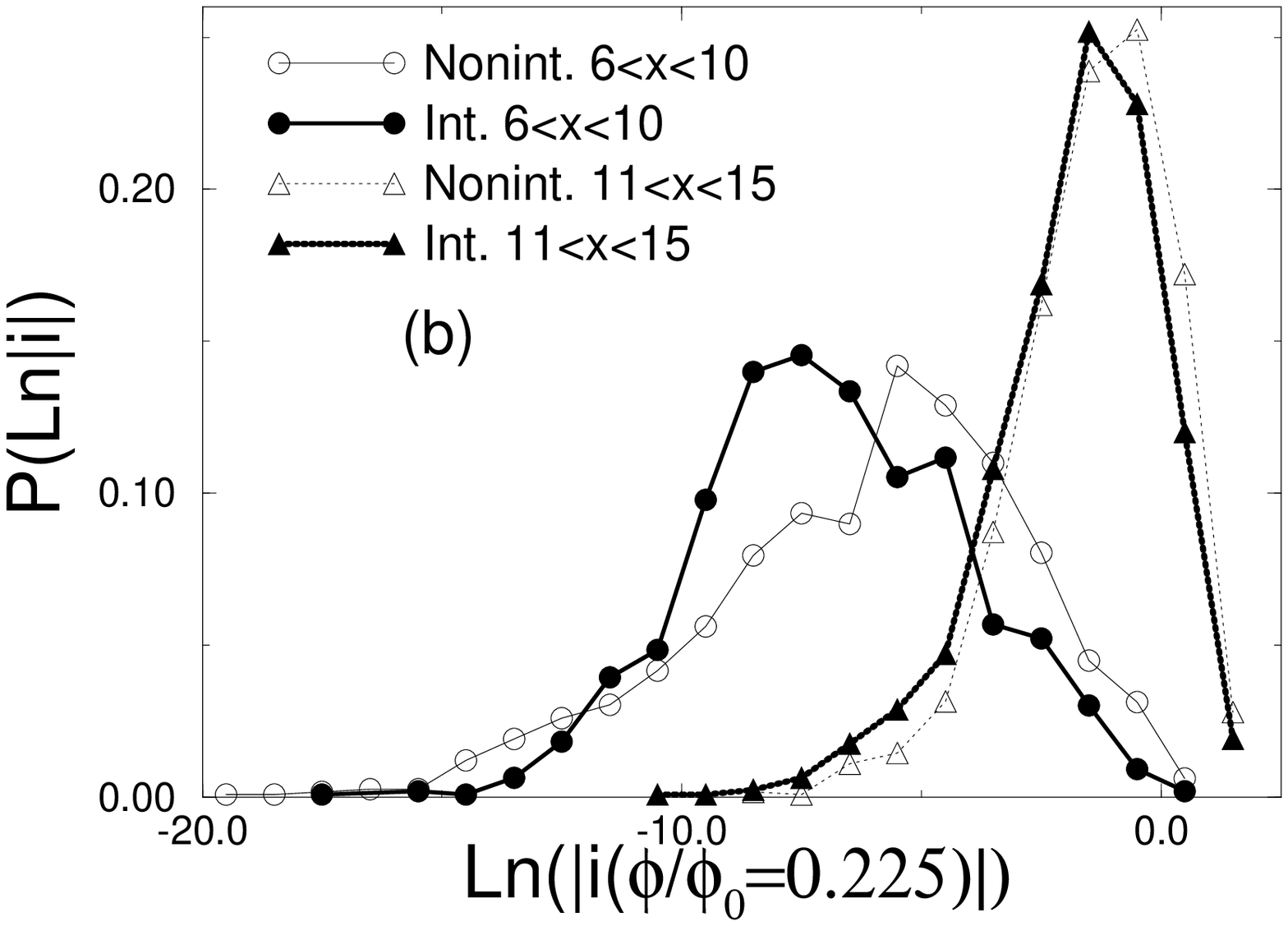}}
\caption{
Probabilities for level PC of 
interacting (bold symbols) and noninteracting
(empty symbols) electrons in the {\em same} canonical ensemble
of 150 disordered rings
at ${\phi \over \phi_0} =0.225$.
(a) For extended levels within two high energy strips.
At the highest strip the states are close to ballistic and
have greater probability to carry large positive or negative current.
At the lower strip the probability is compared with a Gaussian
(average and variance from numerical data)
as predicted by Simons and Altshuler.
Note the interaction induced diamagnetic levels currents.
(b) For localized levels within two low energy strips.
Note the horizontal logarithmic 
scale and the further suppression of PC
due to interaction.
(The visual shift between the distributions
depends on the width of the horizontal bins).
}
\label{fig5}
\end{figure}
To get an information on the single electronic levels
we have also compared the probability distributions of
the single level PC
for noninteracting and interacting electrons.
One should separately analyze the different regimes
of disorder previously defined.
In Fig.\ref{fig5}a the distributions of single level PC are given
for two different high energy strips in the extended regime.
It is evident that due to the interaction the single level PCs are
slightly shifted toward more negative values of the currents. 
The dip at the center of the probability distribution of the levels
belonging to the highest energy strip is due to the fact that the
electrons are almost ballistic and have a greater
probability to carry large positive or negative current. 
In the lower energy strip
the distribution follows a Gaussian shape as was found by Simons and
Altshuler\cite{simons93} for levels in the diffusive regime.         
In contrast to the high energy regime
the effect of the interactions on electrons in
the localized regime is to
suppress the PC. 
This is clearly seen in Fig.\ref{fig5}b where we
considered levels belonging to two different low energy strips
for which the electrons were found to be localized.
Such suppression of the PC in the localized 
regime by e-e interactions was
previously seen in H-F calculations of 
TB models\cite{yoshi}, and in exact
diagonalization of short range interacting electrons\cite{zerar}.
But exact diagonalization studies of TB models with 
long range interaction  
show enhancement in this regime\cite{berkov}. We shall return
to the origin of this discrepancy later on. 

Because the total PC in a sample is 
dominated by electrons at high levels
which have the largest current magnitude
the canonical distribution
of the  sample PC, presented in Fig.\ref{fig4}, is clearly
dominated by the single level
distribution presented in  Fig\ref{fig5}a.
The asymmetry in the PC direction was 
also analyzed in the extended regime
in terms of the distribution of levels curvatures
$k \equiv - {\partial i_l \over \partial (\phi/\phi_0) }$. 
Here the calculations were performed 
by applying the current operator on wave functions because in
this method one can easily calculate the levels PCs at flux values
close to zero or to half flux quantum which, for these flux values,
are proportional to the curvatures\cite{models}.
The distributions of the curvatures were of course similar 
to the appropriate distributions in Fig\ref{fig5}a.

\subsection{Canonical average and enhancement of the single ring PC}

\begin{figure}[t]
\centerline{\epsfxsize = 9cm \epsffile{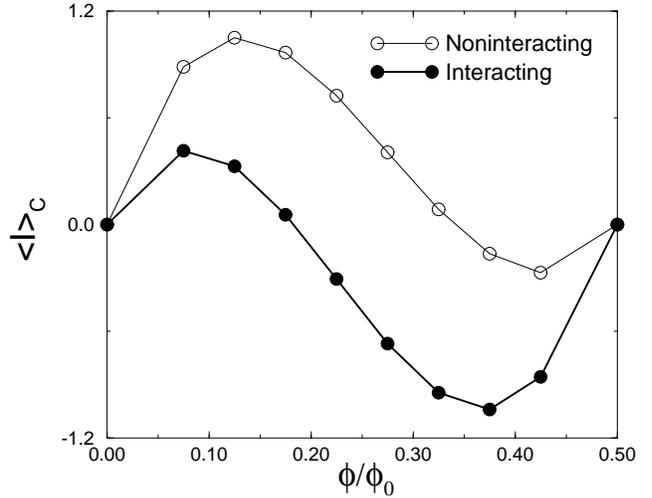}}
\caption{
Comparison of the canonical average persistent 
current of interacting and
noninteracting electrons in an ensemble of 150 disordered rings.
The interaction effect is to shift the noninteracting curve towards
more negative current values when the flux is positive.
The $\phi_0/2$ periodicity is retained.
However it is clear that the amplitude of the average current
is not enhanced.
}
\label{fig6}
\end{figure}

\begin{figure}[t]
\centerline{\epsfxsize = 9cm \epsffile{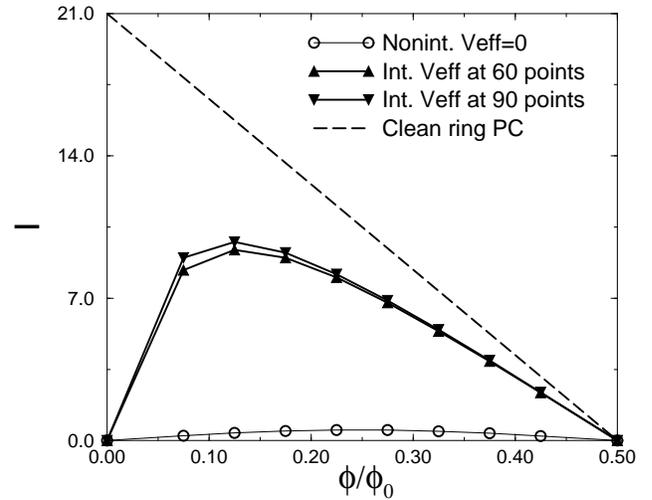}}
\caption{
The highest enhancement factor ($\sim 17$) of single ring 
PC due to the interaction  
that was found among our ensemble. The ring contains 42 electrons.  
The resulting PCs while calculating $V_{eff}(\theta)$ 
at 60 and 90 points are compared. 
This figure demonstrates the possibility of a huge PC  
enhancement found in the single ring experiment.
}
\label{fig7}
\end{figure}

In the previous sub-sections we have presented data on the
distribution of the canonical sample PC, single level PC and single
level curvatures.
We have seen a gradual shift in all of these  distributions
as a consequence of the e-e interactions.
The sum over all electronic contributions from all samples
gives the total current, and divided by the number of samples 
the average current. 
The average current as a function of flux is presented in
Fig.\ref{fig6}.
Though the diamagnetic shift is clearly seen,
the current amplitude is quite unchanged.
This is very different than the suppression found in the various studies 
of 1D TB models in the weakly disordered regime \cite{berkov,zerar,yoshi}.
Nevertheless, this does not present the whole picture.
For a particular sample a strong suppression or enhancement
of the PC amplitude due to the e-e interactions is possible. 
For example, a particular realization shown in Fig.\ref{fig7}
demonstrates the
highest effect of interaction on the PC of a single isolated ring
that we found in our ensemble. The enhancement factor
in this particular
case is about $17$. Of course, even for this case, the
current can not reach its clean ring value. 

\section{conclusions}

Our results within the self consistent HF approximation for the PC 
carried by spinless electrons in 1D disordered continuous rings 
with long range e-e interactions can be summarized 
as follows:  
The PC of states in the diffusive and ballistic
regimes are not suppressed  by the e-e interactions. This is in 
contrast to the suppression of the PC found in 
Refs. \onlinecite{berkov,zerar} and \onlinecite{yoshi} for the 
analogous TB models in these regimes of disorder.   
We found that the PC amplitude for the specific samples may be
altered, but not the amplitude of the average or typical current current.     
Further, the distribution of the PC's acquires a diamagnetic shift.
This was confirmed by comparing the distributions of the individual 
levels currents, and also of the total currents, 
for noninteracting and interacting electrons.
Because the total current of a sample is dominated by the extended 
states the average current per sample also shows a clear diamagnetic behavior.
Our results are significantly 
different from those of TB models where the Mott-Hubbard 
transition was shown to plays a crucial role in suppressing the PC.
In the continuous model the Mott-Hubbard transition does not show 
such a significant role. The concept of band and filling factor does 
not seem to be relevant. 
In the localized regime we found that the PC is further 
suppressed by the e-e interactions. This is an artifact of the
H-F approximation which does not describe the appearance of density
correlations in this regime. An 
exact diagonalization of small 1D TB rings with long-range interaction
show that the PC is actually enhanced in this regime \cite{berkov}.
By studying the effect of e-e interaction on the level spacing
statistics it is possible to gain a better understanding of the
behavior of the current. The level spacing statistics
show no sign of insulator to metal transition and this is 
consistent with the distributions found for the levels PC's.

Introducing spin is expected to allow a stronger 
enhancement of the average current amplitude 
because of the fact that for spinless electrons the  
direct term is almost equal the absolute value of the 
exchange term, while when including the spin degree of freedom 
the direct term is about twice larger. 
This is 
crucial \cite{models} for counteracting the disorder 
potential which leads to a higher PC amplitude.   
This explains the important role of spin for the 
enhancement of the PC as observed for 2D TB 
rings\cite{zerar3da} when compared to the spinless case\cite{zerar3d}.
As we have seen in the 1D case the interactions are effective
in changing the current in the regime of weak-disorder. 
One would expect the interaction to play an important role in 
the diffusive regime. In 1D systems the states are either 
localized or ballistic and therefore, the PC of the states are not 
strongly affected by the interactions.   
On the other hand, 
for higher dimensions where a 
true diffusive regime exists, one might expect for a 
stronger enhancement.

We conclude that the effect of e-e interactions in favor of the  
enhancement of the PC in a continuous disordered model is stronger 
than in TB models. 
This should be better revealed when 
one considers spin and higher dimensionality. 
We have shown that even for interacting spinless electrons in 
1D no further suppression of the PC due to interaction occurs. 
This is in contrast to the further suppression of the current  
in the weakly disordered TB rings that was found in 
Refs.\onlinecite{berkov,zerar} and\onlinecite{yoshi}.

\section*{Acknowledgments}

One of us, A.C., would like to thank B. Shapiro for many valuable
discussions and criticism in the course of this work;
W. Appel, S. Fishman M. Marinov, D. Schmeltzer and S. L. Drechsler 
for valuable discussions in various stages of this work; M. Revsen and
G. Shaviv for computer support at the Technion at the initial stage
of the numerical calculations; B. Kramer and R. W\"oger for
the invitation to P.T.B.-Braunschweig 
and J. Fink for the invitation to
IFW-Dresden that make this work possible.
R.B. would like to thank
the US-Israel Binational Science Foundation
for financial support.



%
%


\begin{references}
\bibitem{lscaling}
The fluctuations flows with $N_s$ towards a universal lognormal
distribution characterized by only two parameters:
the average and variance of $\ln (1+\rho)$ both grow linearly
with $N_s$. The microscopic values at $N_s=1$ are dictated by
the nonuniversal microscopic fluctuations, i.e. the
microscopic disorder.
\bibitem{atvmin} At this point we defined the origin at the
left point of the (sub)interval where $V_{eff}(\theta_k)=0$.
Note also that only when completing a $2\pi$ propagation of the
solution around the ring one has in hand a transfer matrix.
The spectrum and eigen functions are independent of the shift
of the origin.
\end{references}
\end{document}